\documentclass[twocolumn,prl,aps,superbib,tightenlines,floatfix]{revtex4}
\usepackage{amsfonts}
\usepackage{amsmath}
\usepackage{amssymb}
\usepackage{graphicx}

\begin{document}

\title{Role of heating and current-induced forces in the
stability of atomic wires}
\author{Z. Yang\cite{ZY}}
\affiliation{Surface Physics Laboratory (National Key Laboratory), Fudan University,
Shanghai, 200433, China}
\author{M. Chshiev}
\affiliation{Department of Physics, Virginia Polytechnic Institute and State University,
Blacksburg, Virginia 24061-0435}
\author{M. Zwolak}
\affiliation{Physics Department, California Institute of Technology, Pasadena, California
91125}
\author{Y.-C. Chen}
\affiliation{Department of Physics, University of California, San Diego, La Jolla, CA
92093-0319}
\author{M. Di Ventra\cite{MD}}
\affiliation{Department of Physics, University of California, San Diego, 9500 Gilman
Drive, La Jolla, CA 92093-0319}

\begin{abstract}
We investigate the role
of local heating and forces on ions in the stability of
current-carrying aluminum wires. We find that heating increases
with wire length due to a red shift of the frequency spectrum.
Nevertheless, the local temperature of the wire is relatively
low for a wide range of biases provided good thermal contact
exists between the wire and the bulk electrodes. On the contrary,
current-induced forces increase substantially as a function of
bias and reach bond-breaking values at about 1~V. These
results suggest that local heating promotes low-bias instabilities
if dissipation into the bulk electrodes is not efficient, while
current-induced forces are mainly responsible for the wire
break-up at large biases. We compare these results to experimental
observations.
\end{abstract}

\maketitle

\input epsf.sty \flushbottom %\topmargin -20mm

Atomic wires are an ideal testbed to study transport properties at the nanoscale~\cite{muller,agrait}. 
Physical phenomena that have been investigated in these systems include their quantized 
conductance~\cite{cuev}, conductance and noise oscillations as a function of the wire
length~\cite{langPRL,thyg,smit,chendiventra}, 
heating~\cite{agrait,chendiventraNanoLett,todorov} and current-induced atomic 
motion~\cite{yang,todorov,rubio2,diventratodo}. The latter two properties, in
particular, are of key importance in understanding the stability
of atomic wires under current flow~\cite{diventratodo}. 

Halbritter~\textit{et. al.}~\cite{halbritter} and, more recently, Mizobata~%
\textit{et. al.}~\cite{mizo,mizo1} have explored the mechanical
stability of atom-sized Al wires. These
authors found that the probability of forming single-atom
contacts decreases with increasing bias and that it vanishes at a
critical bias of about 1~V~\cite{yan,yasuda}. In addition, several samples 
exhibited low-bias instabilities (typically at biases of 100~mV or less) leading to
break-up of the atomic wires. Similar
trends have been reported in the case of Pb~\cite{halbritter} and
Au~\cite{smit} point contacts, suggesting that the mechanisms 
leading to such instabilities are material independent. 
These low-bias instabilities have typically been attributed to
inadequate dissipation of heat into the bulk 
electrodes.~\cite{todorov,chendiventraNanoLett,smit} However, additional forces on ions due
to current flow may also contribute, since both effects are
present at any given bias. 

In this Letter, we explore the relative role of local heating and 
current-induced forces in the stability of aluminum wires at different biases. 
We study these effects perturbatively, i.e., we first
calculate current-induced forces on ions while neglecting local heating;
second, we assume current-induced forces are negligible, and evaluate the
local temperature of the wires. This perturbative
approach is supported \textit{a posteriori}: we find that local
heating is substantial at very low biases if poor thermal contacts exist
between the wire and the bulk electrodes~\cite%
{chendiventraNanoLett,todorov,smitnanotech}, while current-induced
forces (in particular, the average force per atom, see below) are
small at low biases but increase substantially with increasing external
voltage. In addition, if the heat in the wire can be dissipated
efficiently into the electrodes, the local temperature in the
junction will be quite low at biases for which 
current-induced forces reach bond-breaking values. These
results suggest that low-bias instabilities of atomic wires, 
if any, are due to local heating while high-bias
break-up of a junction is mainly due to current-induced forces.

\begin{figure}
\includegraphics*[width=6.0cm]{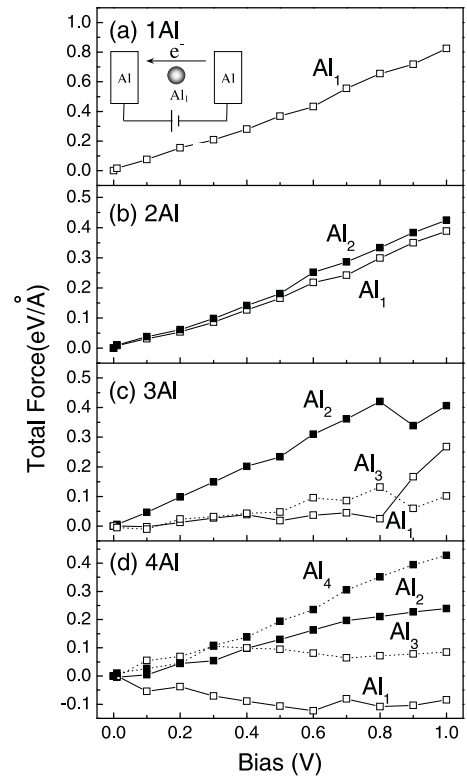}
\caption{Total current-induced forces as a function of bias in
atomic wires containing (a) one, (b) two, (c) three, and (d) four
atoms. The inset shows a schematic of one of the wires
investigated. The atoms are labeled from the left
electrode. The left electrode is positively biased. Positive force
pushes the atom against electron flow.} \label{fig1}
\end{figure}

A schematic of one of the wires investigated is depicted in the
inset of Fig.~\ref{fig1}(a). It consists of an Al atomic chain
(the inset shows a one-atom wire) sandwiched between two Al
electrodes that we model with ideal metals (jellium model,
r$_{\text{s}}\approx $ 2)~\cite{lang1,lang2,diventra1}. The
stationary scattering wave functions of the whole system are
calculated by solving the Lippmann-Schwinger equation iteratively
to self-consistency. Exchange and correlation are
included in
the density-functional formalism within the local-density approximation~\cite%
{lang1,lang2,diventra1}. More details of the calculations of current, forces
and heating can be found in the original papers~\cite%
{diventra1,diventra2,chendiventraNanoLett,diventratodo}.

We investigate the transport properties of Al chains with up to four atoms. All
atomic coordinates have been relaxed to the equilibrium value at zero bias~%
\cite{note1}. The relaxed Al-jellium edge bond length is about 2.0 a.u. and
the relaxed Al-Al bond distance decreases from 6.8 a.u. in the two-Al wire
to about 5.8 a.u. in the four-Al wire. The bond lengths in longer wires are
shorter and approach the bulk bond length, due to the reduced interaction of
the middle atoms with the electrodes.

\textit{Current-induced forces -} The force on an ion, with position $%
\mathbf{R}$, due to current flow is calculated as~\cite%
{diventra2,tod01,diventratodo}
\begin{equation}
\mathbf{F}=-\sum_{i}\langle \psi _{i}\left\vert \frac{\partial v}{\partial
\mathbf{R}}\right\vert \psi _{i}\rangle -\lim_{{\Delta }\rightarrow
0}\int_{\sigma }dE\langle \psi _{\Delta }\left\vert \frac{\partial v}{%
\partial \mathbf{R}}\right\vert \psi _{\Delta }\rangle   \label{equation1}
\end{equation}%
where $v$ is the electron-ion interaction. The first term on the RHS of Eq.~(%
\ref{equation1}) is similar to the usual Hellmann-Feynman
contribution to the force due to localized electronic states
$|\psi _{i}\rangle $. The second term is the contribution due to
continuum states~\cite{diventra2}. The wavefunctions $|\psi
_{\Delta }\rangle $ are eigendifferentials for each energy
interval $\Delta $ in the continuum $\sigma $~\cite{diventra2}.
Finally, the forces due to ion-ion interactions are added to obtain
the total force.
The total current-induced forces as a function of applied
bias in Al wires containing one, two, three and four atoms are shown in
Fig.~\ref{fig1}(a), (b), (c), and (d), respectively. The
atoms in the chains are denoted as Al$_i$ ($i = 1,\ldots,4$) starting from
the left electrode. Positive force indicates atoms are pushed to
the right, i.e., against the electron flow. For the small external
bias of 0.01 V, the forces on Al$_{1}$ and Al$_{2}$ in the
2-Al wire are in the same direction and have almost equal values (see Fig.~%
\ref{fig1}(b)), indicating the symmetry of the two atoms. For 3-Al and 4-Al
wires, the forces obey the zero-sum rule as also found in Refs.~\onlinecite%
{todorov2,yang}: the force on Al$_{2}$ is opposite to that of Al$_{1}$ and Al%
$_{3}$ in the 3-Al wire, and the ones on Al$_{1}$ and Al$_{4}$ are opposite
to those of Al$_{2}$ and Al$_{3}$ in the 4-Al wire, and the sum of the
forces of each wire is close to zero.

\begin{figure}
\includegraphics*[width=6.0cm]{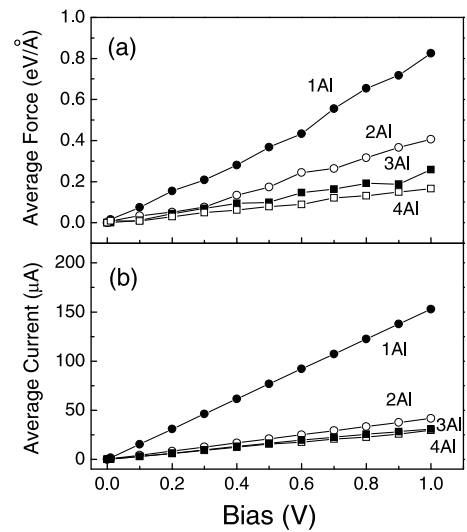}
\caption{(a) Average force on ions as a function of bias for Al wires of
different length. (b) Average current per ion as a function of bias for the
same wires.}
\label{fig2}
\end{figure}

For 1-Al and 2-Al wires current-induced forces move the atoms against the
electron flow at any bias (see Fig.~\ref{fig1}) and increase quite linearly
with increasing voltage up to 1V. On the other hand, for the 3-Al and 4-Al
wire cases, the forces on ions are nonlinear at small voltages with some
atoms moving against and some along with the current flow.
The direction and magnitude of these forces is related to the local resistivity dipoles
that form around the atoms~\cite{yang}.

Despite these nonlinearities as a function of bias, the average force per
atom follows a simple trend: it increases almost linearly with bias and for
a given external voltage it is smaller the longer the wire (see Fig.~\ref%
{fig2}(a)). Similar results have been reported for Si atomic chains~\cite%
{yang}. This trend correlates well with corresponding
current per atom as a function of bias (Fig.~\ref{fig2}(b)) and it
shows that both the force and current per atom saturate with
length. This is consistent with the fact that (i) the wire has no
defects and the resistance is mainly determined by the contacts,
and (ii) the average force due to charge redistribution reaches a 
constant value with increasing wire length, since the stationary states  
become evenly distributed across the length of wire.

While it is difficult to predict the exact bias at which an atomic
chain breaks due to current-induced forces, we can estimate it
using total-energy calculations. For Al atomic wires, Jel\'{\i}nek
\textit{et al.} have found by first-principles calculations that
the maximum force the wires can stand is about 0.6 eV/\AA
~\cite{jelinek}. From Fig.~\ref{fig2}(a) such force
corresponds to a bias of about 0.8~V. This value compares very
well with the reported experimental result of 0.8~V by Mizobata
\textit{et. al.} for single-atom Al contacts~\cite{mizo,mizo1}.
It is also evident that, on average and everything else
being equal, larger voltages are required to break longer wires
(Fig.~\ref{fig2}(a)). However, the break-up of the wires at high bias is likely to 
nucleate from the bonds of the atom with the largest force. 

\textit{Local heating -} We now explore the
effect of heating assuming current-induced forces are zero.
Heating occurs when electrons exchange energy with the lattice via
absorption and emission of vibrational modes. Details of the
theory of local heating in nanoscale structures can be found in
Refs.~\onlinecite{chendiventraNanoLett,todorovjpc}. Here we just mention
that there are two major processes that lead to a given local
temperature in a nanojunction. One is due to inelastic processes
that occur in the region of the junction. In this case electrons
incident from the right or left electrode can absorb (cooling) or
emit (heating) energy because of electron-vibration scattering in
the junction. The other is due to dissipation of energy into the
bulk electrodes. Let us first focus on the inelastic scattering contribution
assuming the dissipation into the electrodes is negligible. This can be the result
of, e.g., weak coupling of vibrational modes localized in the
junction with the continuum of modes of the bulk
electrodes~\cite{todorovjpc}. 

\begin{figure}
\includegraphics*[width=6.0cm]{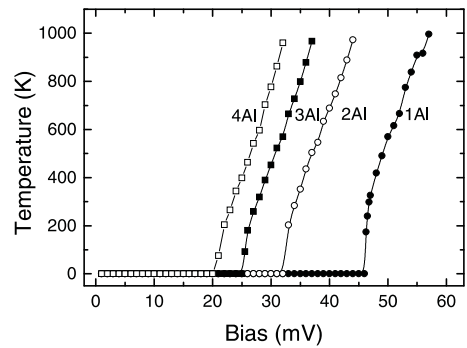}
\caption{Local temperature as a function of bias for Al wires of different
lengths. No heat dissipation into the bulk electrodes is taken into account.}
\label{fig3}
\end{figure}

Denoting by $W_{\nu}^{L(R),1(2)}$ the power absorbed (emitted) by electrons incident
from the left (right) via a vibrational mode $\nu$, the total
thermal power generated in the junction can be written as the sum
over all vibrational modes of four scattering
processes~\cite{chendiventraNanoLett}:
\begin{equation}
P=\sum_{\nu \in vib.}\left( W_{\nu }^{R,2}+W_{\nu }^{L,2}-W_{\nu
}^{R,1}-W_{\nu }^{L,1}\right)  \label{power}
\end{equation}
This power can be expressed in terms of
the electron-vibrational coupling in the presence of current.
This coupling, in turn, can be determined from the same stationary
scattering wave functions used to calculate forces on the 
ions~\cite{chendiventraNanoLett}. When the heating processes
($W_{\nu }^{R,2}$\ and $W_{\nu }^{L,2}$) balance the cooling
processes ($W_{\nu }^{R,1}\ $and $W_{\nu }^{L,1}$), i.e., $P=0$, a
steady-state local temperature is established in the junction.
This temperature is plotted in Fig.~\ref{fig3} assuming zero
background temperature.

A bias larger than the first vibrational frequency of the junction
is necessary to generate heat (see Fig.~\ref{fig3})~\cite{prec}.
One can see that the equilibrium temperature increases
abruptly above that threshold bias and becomes substantial at
biases of only few mV (Fig.~\ref{fig3}). This implies
that if the heat generated in the junction is not efficiently
dissipated into the electrodes, very large local temperatures will be 
generated even at low biases; thus inducing structural instabilities
and consequent wire break-up. These instabilities have indeed
been observed
in Al wires~\cite{halbritter,mizo,mizo1}, as well as in other metallic wires~%
\cite{smit,fujii}. Due to a red shift of the
frequency spectrum with increase of wire length at a given bias,
longer wires have substantially larger local temperatures. Note
that at these same biases, current-induced forces are
negligible (see Fig.~\ref{fig1}).
\begin{figure}
\includegraphics*[width=6.0cm]{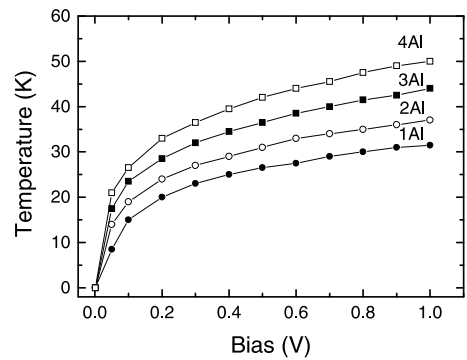}
\caption{Local temperature as a function of bias for Al wires of different
lengths. Heat dissipation into the bulk electrodes is taken into account.}
\label{fig4}
\end{figure}
However, if there is sufficient coupling between the junction vibrational modes and 
the modes of the bulk electrodes, this will allow the heat to dissipate from the
junction into the bulk electrodes via elastic phonon scattering~\cite%
{chendiventraNanoLett,precheat}. This results in a much lower 
local temperature of the junction (see Fig.~\ref{fig4}). It is
clear that at the biases where current-induced forces reach
bond-breaking values, the temperature in the junction is still
quite low (less than about 50~K for all wires at 1~V) with shorter
wires having lower temperatures.

In conclusion, we find that both local heating and current-induced forces can lead 
to structural instabilities, but at different ranges of the external bias. If there is 
poor thermal contact between the junction and the electrodes, local heating induces 
instabilities in the structure at small biases. At high-biases, 
break-up of nanoscale junctions is mainly due to current-induced forces. 
These results help illuminate a very important aspect of nanoscale electronics.

ZY acknowledges support from the NSF Grant No. 10304002 of China. MD
acknowledges support from the NSF Grant No. DMR-01-33075. MZ 
acknowledges support from an NSF Graduate Fellowship.


\begin{thebibliography}{**}
\bibitem{note1} The initial Al-Al and Al-jellium edge distances have been
chosen to be 5.4 a.u. and 2.6 a.u., respectively, see Ref.~\onlinecite%
{lang1}.
\end{thebibliography}
\end{document}